\documentclass[sigconf]{acmart}

\renewcommand\footnotetextcopyrightpermission[1]{}

\AtBeginDocument{%
  \providecommand\BibTeX{{%
    \normalfont B\kern-0.5em{\scshape i\kern-0.25em b}\kern-0.8em\TeX}}}

\usepackage{tikz}
\usepackage{algorithm}
\usepackage{algorithmic}
\usepackage{graphicx}
\usepackage{textcomp}
\usepackage{xcolor}
\usepackage{multirow}
\usepackage{hyperref}
\usepackage{thumbpdf}
\usepackage{booktabs}
\usepackage{colortbl}
\usepackage{url}
\usepackage{subfigure}
\usepackage{array}
\usepackage[normalem]{ulem}
\usepackage{xspace}
\usepackage{wrapfig}
\usepackage{tcolorbox}
\usepackage[framemethod=tikz]{mdframed}
\usepackage{enumitem}

\usepackage{fontawesome}

\def\ie{\textit{i.e.},~}
\def\etal{\textit{et al.}~}
\def\etc{\textit{etc.}~}
\def\eg{\textit{e.g.},~}

\def\Snospace~{\S{}}

\renewcommand*{\figureautorefname}{Figure}

\usepackage{makecell}

\newcommand\boxwidth{8.5cm}
\newcommand\innerwidth{2mm}

\newcommand{\figref}[1]{\figureautorefname~\ref{#1}}
\newcommand{\secref}[1]{Section~\ref{#1}}
\newcommand{\tabref}[1]{Table~\ref{#1}}

\newcounter{insightC}

\newcommand{\insight}[5]{
\refstepcounter{insightC}
\begin{center}
\begin{tcolorbox}
[colback=gray!10,%
  colframe=gray!10,%
  width=\boxwidth,%
  arc=2mm, auto outer arc,
  boxrule=0.5pt,
  left=\innerwidth,
  right=\innerwidth,
]      
\textbf{Incident \arabic{insightC} Title:~}{#1} 

\textbf{Description:~}{#2}

\textbf{Severity:~}{#3}

\textbf{Start time:~}{#4}

\textbf{Service:~}{#5}

\end{tcolorbox}
\end{center}}

\newtcolorbox{mybox}[2][]{colbacktitle=red!10!white, colback=gray!10!white,coltitle=gray!70!black, title={#2},fonttitle=\bfseries,#1}

\newcommand\cloud{Microsoft\xspace}
\newcommand\name{\textsc{Oasis}\xspace}

\newcommand{\point}[1]{\vspace{1mm}\noindent\textbf{#1}.}

\newcommand{\new}[1]{{#1}}

\newcommand{\archive}[1]{} 

\newcommand{\vspacefigdown}{\vspace{-10 pt}}
\newcommand{\vspacetbdown}{\vspace{-6 pt}}

\acmConference[ESEC/FSE 2023]{The 31st ACM Joint European Software Engineering Conference and Symposium on the Foundations of Software Engineering}{11 - 17 November, 2023}{San Francisco, USA}



\begin{document}

\title{Assess and Summarize: Improve Outage Understanding with Large Language Models}

\settopmatter{authorsperrow=4}

\author{Pengxiang Jin*\thanks{*Work done mainly during internship at Microsoft Research Asia.}, \\ Shenglin Zhang}
\affiliation{
    \institution{Nankai University}
    \country{China}
}

\author{Minghua Ma}
\affiliation{
    \institution{Microsoft}
    \country{China}
}

\author{Haozhe Li}
\affiliation{
    \institution{Peking University}
    \country{China}
}

\author{Yu Kang, Liqun Li, \\ Yudong Liu, Bo Qiao}
\affiliation{
    \institution{Microsoft}
    \country{China}
}

\author{Chaoyun Zhang, \\ Pu Zhao, Shilin He}
\affiliation{
    \institution{Microsoft}
    \country{China}
}

\author{Federica Sarro}
\affiliation{
    \institution{University College London}
    \country{UK}
}

\author{Yingnong Dang, \\ Saravan Rajmohan}
\affiliation{
    \institution{Microsoft}
    \country{USA}
}

\author{Qingwei Lin, \\ Dongmei Zhang}
\affiliation{
    \institution{Microsoft}
    \country{China}
}

\renewcommand{\authors}{Pengxiang Jin, Shenglin Zhang, Minghua Ma, Haozhe Li, Yu Kang, Liqun Li, Yudong Liu, Bo Qiao, Chaoyun Zhang, Pu Zhao, Shilin He, Federica Sarro, Yingnong Dang, Saravan Rajmohan, Qingwei Lin, Dongmei Zhang}
\renewcommand{\shortauthors}{P.Jin, et.al.}

\begin{abstract}
Cloud systems have become increasingly popular in recent years due to their flexibility and scalability.
Each time cloud computing applications and services hosted on the cloud are affected by a cloud outage, users can experience slow response times, connection issues or total service disruption, resulting in a significant negative business impact.
Outages are usually comprised of several concurring events/source causes, and therefore
understanding the context of outages is a very challenging yet crucial first step toward mitigating and resolving outages.
In current practice, on-call engineers with in-depth domain knowledge, have to manually assess and summarize outages when they happen, which is time-consuming and labor-intensive.
In this paper, we first present a large-scale empirical study investigating the way on-call engineers currently deal with cloud outages at Microsoft, and then present and empirically validate a novel approach (dubbed Oasis) to help the engineers in this task. Oasis is able to automatically assess the impact scope of outages as well as to produce human-readable summarization.
Specifically, Oasis first assesses the impact scope of an outage by aggregating relevant incidents via multiple techniques. Then, it generates a human-readable summary by leveraging fine-tuned large language models like GPT-3.x. The impact assessment component of Oasis was introduced in Microsoft over three years ago, and it is now widely adopted,  while the outage summarization component has been recently introduced, and in this article we present the results of an empirical evaluation we carried out on 18 real-world cloud systems as well as a human-based evaluation with outage owners. The results obtained show that Oasis can effectively and efficiently summarize outages, and lead Microsoft to deploy its first prototype which is currently under experimental adoption by some of the incident teams.
\end{abstract}

\maketitle

\section{Introduction}
\label{sec:intro}
With the trend of large IT enterprises such as Microsoft, Amazon, and Google deploying services to the cloud platforms, cloud systems have had a booming development in recent years \cite{chen2020identifying, chen2019empirical, ma2022empirical}.  
Tremendous efforts have been devoted to improving the reliability of cloud systems, however, unplanned incidents or performance degradation are still inevitable due to the complex and dynamic nature of cloud systems. 
Often these incidents escalate to a so called \textit{outage}, which impacts multiple services and customers.

\begin{figure}[t]
    \centering
    \includegraphics[width=\linewidth]{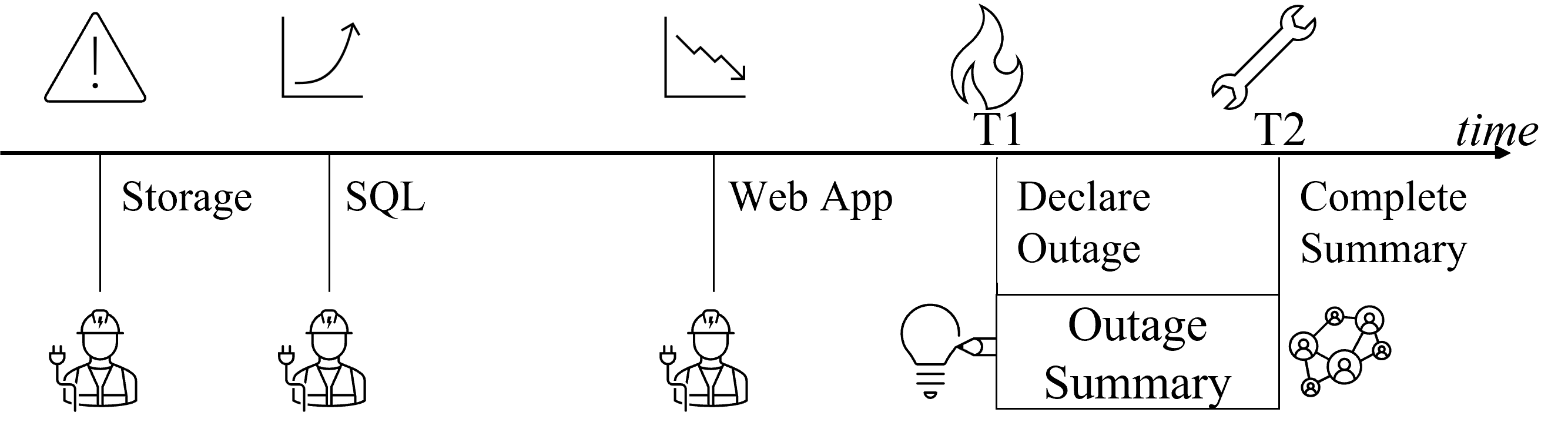}
    \caption{
    The timeline of handling an outage, where multiple incidents should be summarized when declare the outage.  
    }
    \label{fig:timeline}
    \vspacefigdown
\end{figure}

Once an outage occurs to a cloud system, it is crucial to understand its \textit{impact scope} as soon as possible in order to promptly notify customers, mitigate issues \cite{chen2019empirical, ma2020diagnosing}, and ultimately resolve the outage, aiming at reducing as much as possible the loss associate with it.
Nevertheless, a cloud system is quite complex and involves many services such as across-region infrastructures, virtual machines, networking, and database systems, thus making this task very challenging.
To support engineers in monitoring the reliability of the cloud system, each cloud system service has multiple monitors that create an incident each time something wrong occurs. For example, \autoref{fig:timeline} shows the timeline of an incident caused by a flawed configuration change in the Storage service. 
The failed storage affected several SQL databases, and the failure was further propagated to web application instances that depend on the impaired databases. 
Finally, the outage is declared and associated with the multiple incidents occurred in the storage, SQL, and web application services. 
Doing this job manually is not trivial, and in some cases not even feasible. Being able to efficiently aggregate all and only those incidents which are relevant to a given outage, would empower the engineers to promptly investigate the impact scope, as it greatly reduces the number of incidents that need to be investigated.

Previous studies \cite{chen2020identifying,gu2020efficient, chen2021graph, wang2021fast} have devoted a lot of efforts to dealing with incidents aggregation or linking the relevant incidents to the outage. 
However, based on our real-world experience in Microsoft, we observe that on-call engineers (OCEs) still need to manually check the detailed information of relevant incidents  and write a \textit{summary} of outages (a real-world example in \autoref{subsec:rq2}), which is helpful to further handle the outage in terms of notification,  mitigation, diagnosis, and resolution.
To the best of our knowledge, extensive studies on outage understanding are lacking. 
Therefore, in this paper, we first empirically investigate the negative effects of outages in worldwide popular cloud systems in \cloud and how engineers currently deal with them. To this end, we exploit data collected from the usage of 18 real-world cloud systems (many of which are worldwide popular systems) over the past three years.
We found that most outages have a huge negative impact on customers, and the median summarization time is 1 time unit (about one hour)\footnote{Due to the company policy, we hide the actual time and normalize it as time unit.}. 
Therefore, in practice, engineers have to spend significant efforts to understand outages. 
Besides, the content of outage summaries often contains detailed when, where, who, what, and why.  This information is complex and cannot simply adopt as a template because it must be readable by engineers from various component teams.
Thus, it is necessary to automatically summarize outages for understanding quickly. 

These results motivated us to explore automated ways to improve engineers' understanding of outages. To this end, we propose \name, which has two components: \textit{impact scope assessment} and \textit{summary generation}. 
As for impact scope assessment, we adopt three techniques   
\ie rule-based, historical lookup, and deep learning based to aggregate relevant incidents to the outage.
To embed domain knowledge of cloud systems, engineers implement some linking rules from incidents to outages. 
To automatically learn the  correlations among components of cloud systems, we propose a historical lookup algorithm to form a component graph based on the historical incident linkage and match new incidents in the graph.  
To capture the rapid evolution of cloud systems, the deep learning based linking approach is used.
The impact scope of an outage is composed of the relevant incidents aggregated by these three techniques.
We have deployed the impact assessment component of \name in \cloud,
which is running for over three years and achieve significant results in impact scope assessment.

After we obtain relevant incidents of the outage, we adopt the most popular pre-trained large language models GPT-3.x (both GPT-3.0 and GPT-3.5), to automatically generate outage summaries. 
This task presents two main challenges: 1) identifying which information on relevant incidents is helpful to outage summarization; 2) identifying how to effectively generate domain-specific outage summaries with complex cloud-related information. 
For the first challenge, our empirical study provides some guidelines on 
summarizing outages, which reveals the importance of incident severity and description. To tackle the second challenge, we fine-tune the pre-trained large language model, which can generate human-readable sentences and embed with knowledge from cloud systems.   

To investigate the effectiveness of \name, we conduct extensive experiments using real-world outages from \cloud. 
The results show that \name is able to effectively and efficiently generate outage summaries and titles for cloud systems, and significantly outperform all the compared approaches \cite{liu2018neural, radford2019language}. 
More specifically, \name achieves scores of 0.665 (BLEU-4), 0.742 (ROUGE-L), and 0.734 (METEOR) with its summarization 
which outperforms state-of-the-art approaches by at least 32.3\%.
Furthermore, to investigate the usefulness and readability of our generated summaries, we conduct a preliminary human evaluation involving 54 outage owners.  Based on the rankings of summaries produced by models and the original OCEs, we find that \name can achieve human-level summaries much more quickly 
(251.2 times faster than the median of manual summarization). 

Based on the above results, the Oasis outage assessment component has already been in usage for over three years at Microsoft, while the more recent summarization component has been now prototyped and used by some of the incident teams at Microsoft in a phase preceding the final rolling in production.

To sum up, our work has the following contributions: 

\begin{itemize}[leftmargin=*]
\item We are the first to identify outage understanding, a practical scenario for large-scale cloud services.  We have conducted an empirical study of 18 cloud systems to investigate this scenario. 

\item We propose \name, the first automated approach to tackle the problem of outage understanding based on impact scope assessment and large language models (LLMs).  We are the first to propose LLM-based summary generation of outages.

\item Our impact scope assessment of \name has deployed in \cloud for over three years and achieved significant impact. We conduct an extensive study and human evaluation to demonstrate the efficacy and potential usage of \name.
\end{itemize}

\section{Background}
\label{sec:background}
\point{Cloud systems} 
Cloud systems have become increasingly popular in recent years, as they offer a range of benefits such as scalability, accessibility, and cost-efficiency. 
To ensure the reliability of these systems, engineers use various monitoring tools and techniques, \eg Azure Monitor,
to track and analyze the performance and health of different levels and components of the cloud system \cite{chen2020towards, ghosh2022fight}. 
If the monitors detect anomalies, incidents will be reported.

\point{Incidents}
Incidents are unplanned interruptions to cloud service. 
Incident Management is the process of logging those interruptions, and resolving those in a timely manner \cite{chen2019continuous, chen2020incidental, li2021fighting, gu2020efficientincident, jiang2020mitigate, chen2020towards}.
An incident is reported with many fields, for example, the service that the incident is defined on, the source of the incident creation, the time of the incident creation, and a text field describing the problem.
The text description can be generated by the monitor based on pre-defined templates or filled in manually by the engineers.
Moreover, engineers assign a severity level to each incident, ranging from 0 to 4, where a severity of 0 means highest priority and large customer impact, and a severity of 4 means lowest priority.

\point{Outages} 
Outages are severe incidents that require collaboration across many services or result in customer impact \cite{wang2021fast, chen2019outage}. 
Different products and teams may define outages differently depending on service level agreements (SLAs), customer expectations, or other criteria.
When an outage happens, it tends to affect various aspects of the cloud system, causing many incidents to be reported.
OCEs need to go through these incidents to fully understand the outage.

\point{IcM system}
To facilitate mitigating and resolving outages, 
our collaborated \cloud has developed an Incident Management system (IcM) for cloud systems. 
After a monitor reports an incident, an associated incident is created on the IcM.
Then engineers can discuss the incident, check the information, and update the status of the incidents on the IcM page, \etc 
An incident may escalate and is declared as an outage if it impacts multiple services or customers as shown in \autoref{fig:timeline}.
During these processes, records of incidents and logs of the actions are persistently stored in the IcM database.

\section{Outage Understanding: A Case Study}
\label{sec:study}
To better understand the impact of outages and the need for automatic support in outage scope and summary production, we conduct a case study on real-world outages and their summaries.  To this end, we collect outages from 18 systems over three years in the IcM database of \cloud, which serves millions
of daily users worldwide,
specifically, outages and relevant incidents that occurred between January 1, 2020 and October 1, 2022.
To ensure that the outages have undergone careful examination and their summaries are ready, we keep over 6000 outages whose state is `MITIGATED' or `RESOLVED' during collection.
We are not able to make all the details public due to the company's policy. 

In this study, we address the following research questions:

\begin{itemize}[leftmargin=*]
\item \textbf{RQ1}: What is the impact of outages?
\item \textbf{RQ2}: What are the information included in outage summaries?
\item \textbf{RQ3}: What is the cost (in terms of time) of manually summarizing outages?
\end{itemize}

\subsection{RQ1: Impact of Outages}
\label{subsec:rq1}

\point{Impact on customers}
When OCEs deal with outages, it is important to decide the impact on customers, especially the number of customers affected.
For each outage, OCEs determine whether it impacts a large number of users and record this determination. 
We statistically analyze the outages that OCEs considered as impacting a large number of users and found that such outages accounted for as much as 86.4\% of all outages.
Outages usually have a significant impact on cloud systems, resulting in a degraded user experience for a large number of customers.
Therefore, it is crucial to quickly and effectively respond to outages. 

Another aspect of the customer impact is whether an outage has resulted in persistent impacts.
OCEs mark the outages that have persistent or intermittent impacts with a flag variable.   
The number of outages resulting in persistent impacts is 1.81 times 
more than the number of outages that have intermittent impacts.
The impact of an outage on a cloud system is frequently severe.

\begin{figure}[t] 
\centering

\subfigure[Incidents]{
    \includegraphics[width=.45\linewidth]{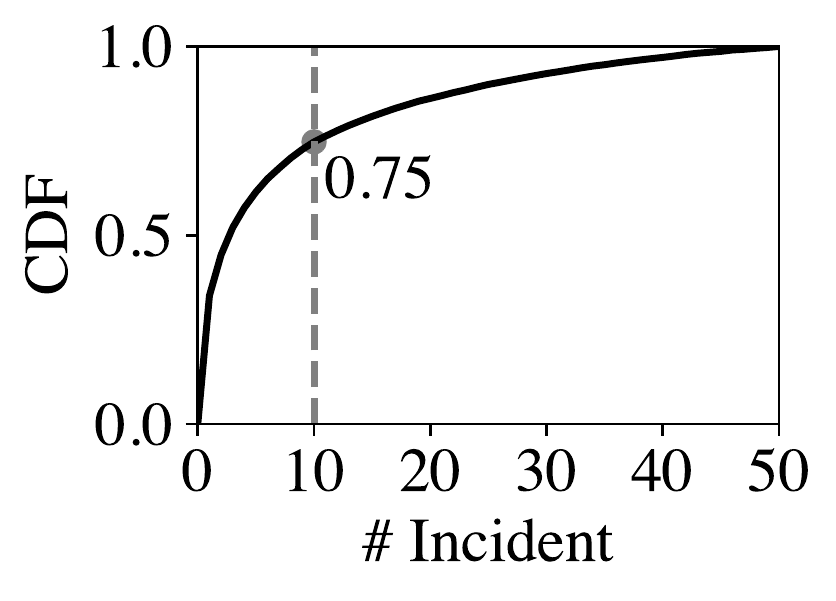}
    \label{fig:relatedincidents}
}
\subfigure[TTS]{
    \includegraphics[width=.45\linewidth]{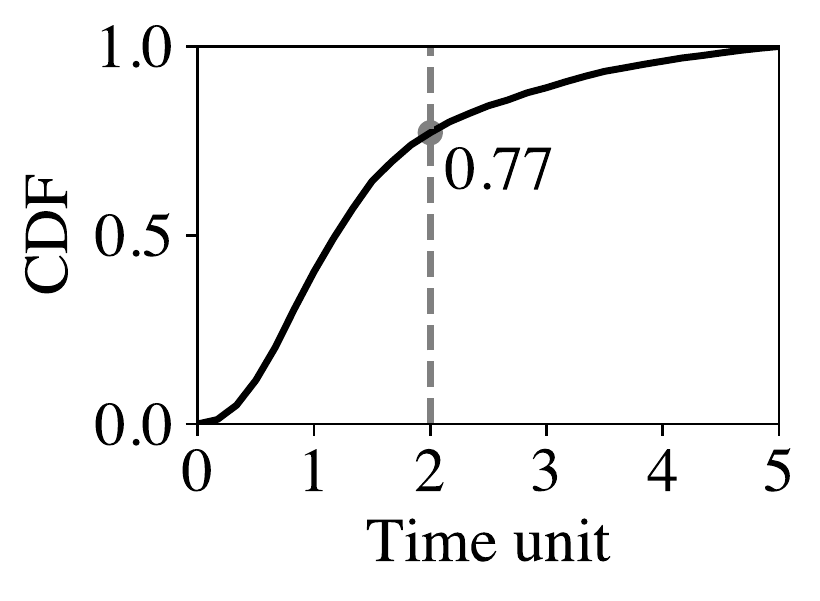}
    \label{fig:tts}
}
\vspace{-13pt}
\caption{
CDF of 
(a) Number of relevant incidents to outages.
(b) Time to Summary (TTS). %
}
\end{figure}

\point{Relevant incidents}
Several incidents in cloud systems are continuously reported and escalate to one outage, as they share a common root cause. 
The distribution of incidents associated with outages is illustrated in \figref{fig:relatedincidents}, with 25\% of outages having more than 10 associated incidents. 
The average number of relevant incidents to an outage is 9.36.
Based on this data, the outages bring about many incidents, consuming the efforts and time of the OCEs.

\subsection{RQ2: Outage Summary Information}
\label{subsec:rq2}

To help understand what information needs to be summarized for an outage,  we demonstrate a real-world outage summary written by OCEs and its relevant incidents.
We mask several details due to confidentiality.

\insight
{Alert: email-api-batchevents-errors-production-allregions-exceeded}
{
    \dashuline{The Email Service was experiencing connec- tivity issues to their replica database in the West US Region. }
    Due to this issue, System-Cloud customers globally were not receiving any type of System-Cloud notifications.
} 
{2}
{14:28 }
{SQL}

 \insight
 {No Success Signal in the last 60 minutes.}
 {
     Calls to the API-Sub failed with a 5xx HTTP error. 
     \dashuline{Approximately $\alpha_1$ customers could not upgrade their subscriptions on URL-Cloud-Portal.}
 }
 {2}
 {15:30 }
 {Commercial}

\insight
{System-Cloud Email Orchestrator Health in Cluster$_1$}
{
	Email notifications sent to customers could be delayed.
 }
 {3}
 {14:33 }
 {Business Intelligence}

\insight
{Api request failed with multiple -1 responses. Target: URL-Cloud-Email}
{
	Calls to the API-Marketplace service failed which prevented the service from sending emails to the customers and affected $\alpha_2$ customers.
 }
 {3}
 {17:06 }
 {Marketplace}

 \insight
 {Email Service calls are failing for Monitor-Email-Exceptions evaluated on MonitorRule$_1$ unhealthy 
 }
 {
	Customers could not view Customer renewal and subscription alerts were delayed. In addition, users were unable to get authentication codes to verify login and new account sign-ups.
 }
 {3}
 {14:44 }
 {Notification}

\noindent
The title, times, and summary of the outage are listed below:

 \begin{center}
\begin{tcolorbox}[colback=gray!10,%
  colframe=gray!10,%
  width=\boxwidth,
  arc=2mm, auto outer arc,
  boxrule=0.5pt,
  left=\innerwidth,
  right=\innerwidth,
]      
\textbf{Outage Title:~} {
    Outage for Email Service - Triage
}

\textbf{Impact start time:} 14:20 

\textbf{Outage declared time:} 14:28 

\textbf{OCEs engage time:} 14:29 

\textbf{Outage Summary:~}{ 
\dashuline{The Email Service experienced connectivity issues to their replica database in the West US Region.}
This affected customer email delivery for approximately $\alpha_3$ internal company services. Due to this issue, System-Cloud customers were not receiving notifications including purchase, renewal, and monitor alert notifications. The Portal team reported that 
\dashuline{approximately $\alpha_1$ customers were unable to upgrade their subscriptions on URL-Cloud-Portal.}

} 
\end{tcolorbox}
\end{center}

We can see from the above example that each relevant incident describes various aspects of the outage, and the information about an outage fall into many different categories.
For example, West US Region is a physical location, and \{System-Cloud, Email Service, API-Marketplace\} are software components at different layers that are affected, and \{$\alpha_1$, $\alpha_2$, $\alpha_3$\} are specific numbers describing the number of impacted customers or services, and \textit{5xx HTTP error} is a software bug that affects the service functionality.
Formally, the information of an outage usually involves 5W (when, where, who, what, why): 

\point{When}
\textit{When does the outage start impact, get declared, and engaged?} 
Engineers pay attention to several time points and periods of an outage.
For example, the time when the outage starts to make an impact, when the outage is declared, and when the OCEs start to engage are important signals for assessing the availability and reliability of the system.
Additionally, when assessing the impact of an outage, it is also useful to know the time window period when a certain function is unavailable.

\point{Where}
\textit{Where does the outage come from?}
The physical location of an outage can lie in various levels of the cloud infrastructure. The physical location can have an impact on the time required to resolve it and the potential for cascading failures. Additionally, the physical location of an outage can be a key factor in determining the impact on customers, as local or nearby customers may be affected more severely.
The physical location of the \textit{cloud} infrastructure at \cloud is structured in a hierarchical manner \cite{ori2020protean} with \textit{regions} and \textit{availability zones} at the top level, which is directly accessible to customers. Each region can consist of up to three \textit{availability zones}, each containing one or more \textit{datacenters}. These \textit{datacenters} are further divided into \textit{clusters}. 
Despite the fact that other cloud systems may exhibit different location hierarchies, it is as important to know the location of outages.

\point{Who}
\textit{Which services are suffering from the outage?}
Services of cloud systems can be divided into different layers:
(1) application layer: this layer contains the actual code and functionality of the cloud system, where frontend and backend services are located,
(2) platform layer: this layer provides the operating system, middleware, and runtime environment for the cloud system, which may include virtualization software, container orchestration software, or serverless computing framework,
(3) data layer: this layer handles the storage and management of data used by the cloud system, which may include databases, data lakes, and data warehouses,
(4) infrastructure layer: this layer provides the underlying physical and virtual resources that are used to run the cloud system, which includes host servers, storage, and networking.
Each layer has its fine-grained components.
Assessing which parts of the cloud system are affected by the outage is helpful to handle the outages.

\point{What} 
\textit{What happens to the cloud system in the outage?}
Previous research has shown \cite{ghosh2022fight} some common symptoms of outages, including:
(1) \textit{code bugs}, such as buggy or incompatible code that generates error results,
(2) \textit{dependency failures}, such as an unhealthy dependent service that impacts the functioning of downstream services,
(3) \textit{infrastructure issues}, such as high CPU utilization of a server that prevents the service from functioning normally,
(4) \textit{deployment errors}, such as an engineer deploying an incorrect certificate.
There are also other less frequent symptoms, such as \textit{configuration bugs}, \textit{database/network} issues, \textit{authentication failures}, \etc

\point{Why}
\textit{Why did the outage happen?}
Previous research has investigated the four most common root causes of outages \cite{liu2019bugs}:
(1) insufficient or erroneous mechanism of \textit{fault handling} (\eg error component, unresponsive component, and silent corruption),
(2) \textit{data format} incompatibility between different software components,
(3) \textit{timing}(\eg concurrent) bugs,
and (4) misconfigured or outdated \textit{constant values}.
However, the bugs in production cloud systems are highly diversified.
The underlying causes of misbehavior require thorough manual investigation by OCEs.

\point{Focus of outage summarization}
We can see from the example that the summary of an outage is not simply a list of information.
OCEs when writing outage summaries are more favorable to high-severity incidents.
Moreover, textual description is an important reference in the outage summarization process.
For example, the dashed sentences are taken directly from the textual descriptions of two high-severity incidents, \ie Incident 1 and 2.

\noindent
\textit{\underline{Finding}}:
High-severity incidents and their textual descriptions are important in outage summarization.

\subsection{RQ3: Time to Summary}
After an outage starts to make an impact on customers, OCEs need to quickly respond to the outage. One key step is to summarize the context of the outage. Therefore, we investigate the time to manually write outage summaries. Specifically, we retrieved the impact start time  (T1 in \figref{fig:timeline}) of the outage and its summary completed time (T2).
The time needed to summarize outages is calculated by T2 $-$ T1.

\figref{fig:tts} shows the CDF of the time needed to summarize outages. %
From this figure, there are nearly 23\% of outages cannot be summarized within two time units after the outage starts. 
The median time needed to summarize outages is one time unit.
Therefore, outage summarization is time-consuming and labor-intensive.

\subsection{Summary}

According to our empirical study, the impact summary of an outage may include domain specific terminology of physical or logical locations, service name, code change name, \etc 
Besides, what and why of outages are even more difficult to summarize simply using templates.  
OCEs must have a thorough grasp of the relevant incidents in order to effectively summarize an outage, and the text descriptions of these incidents provide crucial information for this comprehension.
Nowadays, pretrained large language models (\eg GPT-x) show their ability in many tasks, such as Q\&A and summarization in ChatGPT.
Therefore, we aim to employ large language models to help outage summarization.
In this work, we aim to generate outage summaries with the following goals:

\point{Usefulness}
Usefulness measures whether the outage summary contains relevant and valuable information.

\point{Readability} 
Readability measures whether the outage summary read fluently, especially considering the context information of the outage and the affected system.

\noindent\textbf{Reducing TTS} (time to summary).
It is desirable to summarize outage in a short time because it helps improve the overall outage handling process, improve communnication, shorten the lifecycle of outage, and in turn, improve customers' satisfaction.

\section{Our Proposal: OASIS}
\label{sec:approach}
\subsection{Overview}
\begin{figure*}[t]
    \centering
    \includegraphics[width=.8\linewidth]{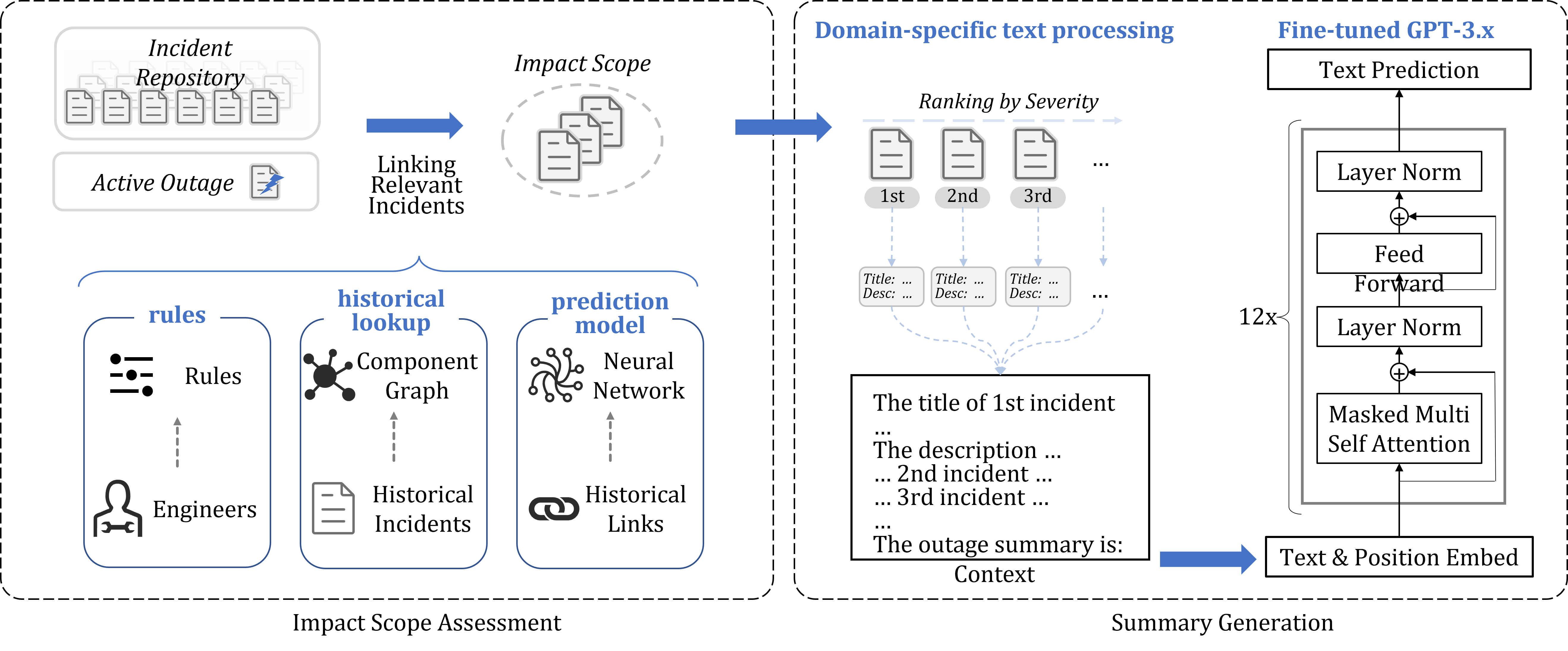}
    \vspacefigdown
    \caption{
        Overview of \name.
    }
    \label{fig:arch}
    \vspacefigdown
\end{figure*}

In this paper, we aim to automatically generate summaries for outages of cloud systems.
However, outage summarization faces two challenges. 
The first challenge is determining which information on relevant incidents is helpful to outage summarization.
Since the cloud system is complex and rapidly growing, it is not trivial to extract domain-specific terminologies of incidents.
The second challenge is how to effectively generate human-readable outage summaries with complex cloud-related information.

To solve these challenges, we propose \name.
The overview of \name is shown in \figref{fig:arch}, which consists of the following two components.
In the first component, \ie impact scope assessment, \name identifies relevant incidents via three types of linking to 
comprehensively assess the impact of the outage.
In the second component, \ie summary generation, \name first performs 
domain-specific text processing to denoise and prioritize important information from the obtained relevant incidents, thus addressing the first challenge.
Then \name employs a fine-tuned large language model, \ie GPT-3.x, to understand the incidents and generate a compact summary for the outage, thus addressing the second challenge.

\subsection{Impact Scope Assessment}
\label{subsec:scope}

Assessing the impact scope of an outage is about comprehensively understanding different aspects of an outage such as the when, where, who, what, why, \etc
As shown in \secref{sec:study}, the impact of these aspects of the outage is collectively described by many relevant incidents.
However, there is no simple and direct way to identify the set of relevant incidents, since incidents with the same underlying root cause can have different properties and spread across services.
Meanwhile, if an OCE determines two incidents are highly relevant to each other, she can formally \textit{link} the two incidents together, which is a feature provided by the IcM system.
Linking an incident with relevant incidents reduce the effort of OCEs in many ways, for example, reducing the number of incidents that require manual examination, auto-resolving less severe incident if a more severe linked incident is resolved, \etc

Inspired by the process, \name assesses the impact scope of an outage by linking its relevant incidents.
To completely link the relevant incidents of an outage, \name incorporates domain knowledge and historical linking patterns.
Specifically, \name performs three types of incident linking: linking by rule, linking by historical lookup, and linking by prediction model.

\point{Linking by rules}
Automated incident linking is a capability in IcM that correlates and de-duplicates incidents to reduce alert storms and noise. 
Engineers can set up specific rules to create links between incidents upon various triggers, which represent the domain knowledge of the engineers.
For example, an engineer can set up a rule to have incidents that are triggered by the same KPI anomaly be linked.
These are structural incident links that can be directly queried from the IcM database.
During the operation of cloud systems and the corresponding IcM system, a large number of historical incidents and rule-based links are persistently recorded.
These data are a natural source of labeling for learning, which facilitates the following two types of linking.

\point{Linking by historical lookup}
We propose a heuristic lookup mechanism to utilize the historical links between incidents.
The mechanism consists of an offline phase to memorize the historical linking pattern, and an online phase to apply the patterns to current incidents.
In each phase, we use the field of component that is reported along with the incident.
Components are fine-grained parts of cloud systems that are defined by engineers.
In the offline phase, we build a component linking graph by summing up links between incidents, \ie if incident A and incident B are linked, then we link their component in the component graph.
In the online phase, we check whether there are active incidents (incidents within a short time range) on the linked components.

\point{Linking by prediction model}
Another way to automatically discover the relationship between incidents is by employing deep learning techniques \cite{chen2020identifying, gu2020efficient}.
It has the advantage of being highly automated and can be applied to a large number of incidents. Also, it can be applied to scenarios where new incident detection criteria are created and the engineers have not set up rules and historical links fail to apply because of the lack of historical data.
We train a neural network to predict the link between incidents.
The neural network takes the titles and descriptions of two incidents as input and outputs the similarities between the titles.
If the similarity of two incidents is larger than a threshold, we determine the two incidents are linked.
The neural network has been trained on pairs of incidents to learn the relationships between incident linking and incidents' titles and descriptions.
In the training set, incident pairs that were actually linked by engineers are labeled positive and pairs that were not linked are labeled negative.

\point{Put them together}
\name periodically assesses relevant incidents to the outage by querying the information of incidents within a time window to the outage.
We take advantage of three linking approaches: the rule-based linking has the highest confidence and interoperability;
the historical lookup may find hidden dependencies;
the prediction model can adapt to the rapid evolution of cloud systems.
Together, these three linking approaches link an outage to a set of relevant incidents, which will be used to generate the outage summary.

\subsection{Summary Generation}

After gathering relevant incidents of an outage, \name generates a summary of the outage based on the incidents' information.
To overcome the challenge of noisy information, we fine-tune an LLM, \ie GPT-x, to summarize the relevant incidents.

\point{GPT-x}
Generative Pre-trained Transformer x (GPT-x) \cite{brown2020language} is a large pretrained language model that can tackle a wide range of natural language processing (NLP) tasks. 
One typical usage scenario of GPT-x is text completion, where the model is given a block of text as \textit{context} and generates text as the \textit{completion} of the context.
It has been explored to recommend the root causes of cloud incidents \cite{ahmed2023recommending}.

The GPT-x model is based on Transformers \cite{vaswani2017attention}, which takes advantage of the attention mechanism to assign weights to different parts of the text.
Thus it is suitable to summarize noisy information.
There are different sizes (number of parameters) of GPT-x model.
In our work, we implement \name with two parameter sizes: \textit{GPT-3-Curie} and \textit{GPT-3.5-DaVinci} (see \secref{sec:experiment-environment} for more details).

\point{Domain-specific text processing}
In this step, we process the structural incident information into a paragraph of text so that the GPT-3 model can take it as input, \ie context.
Inspired by the findings from \secref{sec:study}, we propose to process incidents in a way that the high-severity incidents and textual descriptions are emphasized.
First, we sort the relevant incidents by their severity so that the incidents with higher severity precede the ones with lower severity.
Then we transform the incidents into a piece of text in the following way:
for the incident with sorted order $i$, the text is [\textit{
    The title of $i^{th}$  incident is \dots.
    The description of $i^{th}$ incident is \dots.
    The service of $i^{th}$ incidents is \dots.
    \dots
}]
Finally, we append an instruction to the end of the text to hint the GPT-3 model to generate a summary: [\textit{The outage summary is:}].

\point{Fine-tuning}
The GPT-3 model was trained on a general corpus that allows the model the learn various knowledge like linguistics, common knowledge, factual knowledge, basic logical inference ability, \etc 
To achieve better summary generation, we use our IcM-specific data to fine-tune the GPT-3 model so that it learns the domain knowledge of the applied cloud system and incidents. 
Moreover, the training samples presented to the model teach it to emphasize the aspects that are of interest to OCEs, thereby improving its ability to summarize information from noisy sources.
The data we use to fine-tune the model is in the same form of summary generation, \ie for each outage, we provide the relevant incidents as context and the outage summary written by engineers as the desired completion.

\subsection{System Implementation}
\label{subsec:imp}
We have deployed \name as an aid to the IcM system of \cloud.
We will introduce the integration of \name with the IcM workflow and the underlying implementation details.

\begin{figure}[t]
    \centering
    \includegraphics[width=.8\linewidth]{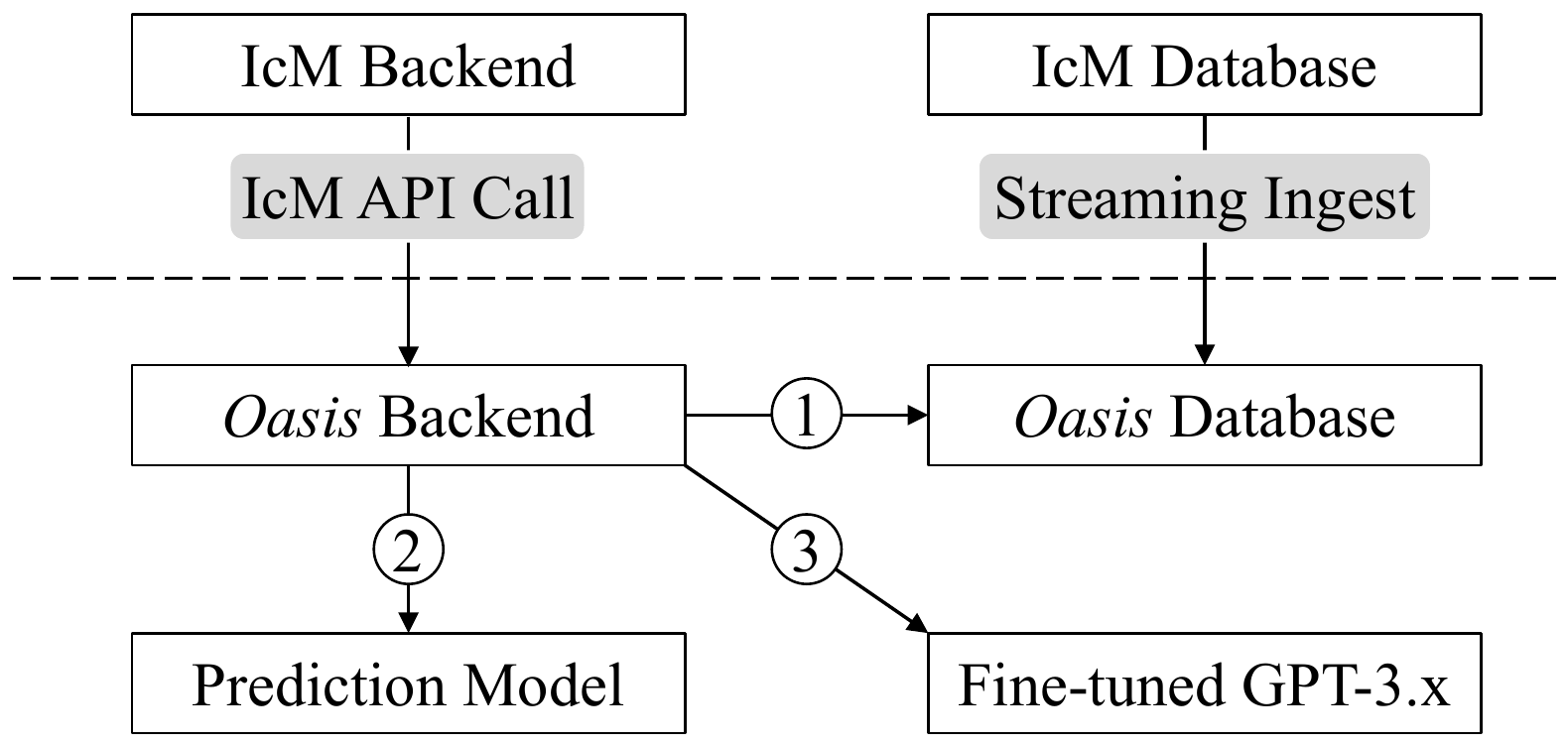}
    \vspacefigdown
    \caption{
        \name in production. The upper part is the IcM system. The lower part is the architecture of \name.
    }
    \label{fig:imp}
\end{figure}

The implementation of \name in production consists of four parts, as shown in \figref{fig:imp}.
The \name backend periodically queries the local database to get active incidents within the time window, as well as rule-based and historical-lookup links of all current outages (1) .
On receiving the API call initiated by the IcM Backend querying a specific outage, the \name backend applies the prediction model to determine what other incidents should be linked to the outage (2).
After that, it performs domain-specific text processing for the outage's relevant incidents and feeds the processed text to the fine-tuned GPT-3.x model (3).
Finally, the backend returns the summary generated by GPT-3.x to the IcM Backend.

The local database of \name is ingested from the IcM database in a streaming manner.
Compared to batch ingestion, streaming ingestion is more smooth in resource utilization.
Moreover, the linking requires real-time data records of incidents.
The prediction model has already been trained in ML-dedicated servers and exported as a binary file to minimize the operation effort of \name in production.

\section{OASIS Evaluation: Empirical Study Design}
\label{sec:experiment}
To assess the effectiveness of OASIS, we investigate the following:
\begin{itemize}[leftmargin=*]
\item \textbf{RQ4}: Is \name effective at summarizing outages?
\\ Generating the summary of outages is the main task of \name. We are interested in the ability of \name to generate a reasonable outage summary with automatic impact scope assessment.
\item \textbf{RQ5}: Is \name effective at proposing outage titles?
\\ The title of an outage is a short, highly abstracted piece of text stating the problem that is happening. Proposing outage titles also demonstrates the ability of \name to understand and summarize the outage.
\item \textbf{RQ6}: Does \name get better at summarizing outages if the outage title is given?
\\ In practical settings, OCEs first write the title of an outage and then write the summary of the outage. We are interested in whether \name can better summarize an outage if the title written by OCEs is also given as part of the context.
\item \textbf{RQ7}: What is the time efficiency of \name?
\\ Since \name needs to work in the production environment, it is important for \name to summarize outages efficiently.
\end{itemize}

\subsection{Study data}
\label{sec:eval-data}

In the study, we applied \name to the same 18 cloud systems and the same time range (3 years) in \secref{subsec:rq1} to evaluate the effectiveness of \name.
\new{In particular, we split the data in chronological order using a 7:1:2 ratio for the training (fine-tuning), validation, and test sets, respectively.
Each data point, representing an outage, is presented as a \textit{context}-\textit{completion} pair.
The \textit{context} consists of the processed text from relevant incidents linked by impact scope assessment.
The \textit{completion}, on the other hand, is provided as the summary of the outage written by OCEs.}

\subsection{Compared approaches}
To better answer the RQ 4 to 7, we compare the performance of \name with some baseline approaches. 
We formulate the task as a text generation problem, therefore we compare with 3 methods that have been proven capable of generating summarization.
In answering each research question, we provide the same context (information of relevant incidents) to the baselines and to the GPT-3 model of \name.
\begin{itemize}[leftmargin=*]
\item Joint incident summary (\textbf{Rule-based}): A straightforward rule-based method that concatenates all the information of incidents. This method imitates the behavior that OCEs read through relevant incidents when handling outages.
\item Information retrieval (\textbf{IR}): NNGen \cite{liu2018neural} leverages bag-of-words embedding and nearest neighbor to retrieve summaries from similar history outages.
\item \textbf{GPT-2}: Generative Pre-training Transformer 2 (GPT-2) is a language model that is trained to generate coherent text. We use GPT-2 with 117M parameters. %
\end{itemize}

\subsection{Metrics}
Following the existing work \cite{ahmed2023recommending, gros2020code, ahmed2022multilingual}, we use the BLEU-4, ROUGE-L, and METEOR to evaluate \name and its baselines in terms of readability. 
The BLEU-4 compares the matching of n-grams between generated text and the ground truth.
The ROUGE-L is widely used in Machine Translation evaluation, which measures the overlap of the longest sequence between hypothesis and reference. 
The METEOR calculates the harmonic mean of unigram precision and recall with consideration of stemming and synonym matching.

Specifically, we get five candidate generated texts from each model, except for the joint incident summary which can only give one piece of generation.
To better evaluate the quality of generation models, %
we calculate the Top1 metrics using the first generated text, and the Top5 metrics using the best of five generated text.

We also measure the running time of each approach.
Specifically, we record the overall time needed to train/fine-tune the model, and the average time spent on generating a summary for an outage.

To further evaluate usefulness and readability, we conduct a human evaluation in \autoref{sec:human}. 
When summarizing outages, the style of OCEs can vary from generic to specific.
Automatic metrics only compare the models' suggestions with a single reference, while other versions of the summary can be useful and relevant as well, so these metrics may not fully capture the performance of models.
To better evaluate the model’s performance, we went to the owners (responsible engineers) of the outages and presented the outputs of our models and baselines.
We will discuss our methodology and findings from the human evaluation in \secref{sec:human}.

\subsection{Experiment environment}
\label{sec:experiment-environment}

\point{Generation model} 
We implement \name with two GPT-3 variants, \ie Curie and DaVinci:
\\ \noindent
\underline{\textit{Curie (GPT-3)}} is a fast GPT-3 model with 6.7 billion parameters, which was pre-trained on a natural language corpus.
\\ \noindent
\underline{\textit{DaVinci (GPT-3.5)}} is a large GPT-3 model with 175 billion parameters, which was pre-trained on both text and code.
\\ \noindent
\new{We fine-tune these generation models using the training and validation set from \autoref{sec:eval-data}.}

\point{Experiment environment} We implement all training with one NVIDIA GeForce A100 GPU, PyTorch 1.11, and CUDA toolkit 11.3.1.

\point{Implementation of baselines} Baselines are implemented using Python 3.8 and scikit-learn 1.0.2. The number of GPT-2's training epochs is 20. The temperature is GPT-2 is 0.7, which is recommended by a previous study \cite{ahmed2023recommending}.

\section{OASIS Evaluation: Empirical Study Results}
\label{sec:evaluation}

\subsection{RQ4: Performance of Summary Generation}

\begin{table}[t]
\caption{
    Effectiveness of models at summarizing outages
}
\begin{tabular}{lllllll}
\toprule
\multicolumn{1}{c}{\multirow{2}{*}{Model}} & \multicolumn{2}{c}{BLEU-4}        & \multicolumn{2}{c}{ROUGE-L}     & \multicolumn{2}{c}{METEOR}      \\
\multicolumn{1}{c}{}                       & Top1           & Top5           & Top1           & Top5           & Top1           & Top5           \\ \hline
IR                                         & 0.042          & 0.051          & 0.144          & 0.180          & 0.115          & 0.146          \\
Rule-based                                 & 0.277          & NA             & 0.508          & NA             & 0.629          & NA             \\
GPT-2                                       & 0.455          & 0.51           & 0.561          & 0.592          & 0.536          & 0.574          \\
Curie                                      & 0.654          & 0.701          & 0.73           & 0.777          & 0.721          & 0.767          \\
DaVinci                                    & \textbf{0.664} & \textbf{0.706} & \textbf{0.742} & \textbf{0.782} & \textbf{0.734} & \textbf{0.776} \\ \bottomrule
\end{tabular}
\label{tab:exp1}
\vspacetbdown
\end{table}

\tabref{tab:exp1} lists the effectiveness of baselines and \name in summarizing outages.
\name with DaVinci, the largest GPT-3 model, achieves the best metrics with both Top1 and Top5 summary generation.
The advantage of DaVinci over Curie comes from the larger parameter size and the extra code corpus used in pretraining since some incidents contain API names or investigating code.
However, the performance gain of DaVinci over Curie, the fastest GPT-3 model, is modest in both Top1 and Top5 generations.

We observe IR method is especially not suitable for outage summary generation.
The major reason that the scores of IR are poor is that the rapid evolution of cloud systems has resulted in significant changes in the architecture of the systems over time, so similar outages are not likely to appear repeatedly, therefore historically useful summaries fail to depict the new outages.
Although Rule-based summaries have a higher METEOR score than GPT-2, their BLEU-4 score is far lower than that of GPT-2. 
This is because the METEOR score takes into account the precision and recall of the unigram rather than subsequences, resulting in a more lenient evaluation of summaries. 
Rule-based summaries are often too long as the original incidents, which is not helpful for engineers to understand the context of outages.

\subsection{RQ5: Performance of Title Generation}
The title of an outage is a compact description of the outage. The example of an outage title and summary is shown in \secref{subsec:rq2}.
The performance of baselines and \name at summarizing outages in the form of titles 
 are listed in \tabref{tab:exp2}.
By comparing \tabref{tab:exp2} with \tabref{tab:exp1}, we achieve a higher generation score (0.826-0.857 BLEU-4) at generating titles for outages than generating the whole summary (0.654-0.664 BLEU-4).
Similarly, GPT2, which is also a large Transformer based language model, scores higher in summarizing outages in the form of a title than the whole summary.
The ROUGE-L and METEOR score exhibit the same trend for \name and GPT-2.
The reason for this performance improvement lies in the nature of outage title and summary.
The title of outage has a stronger pattern than that of outage summary.
Firstly, a large portion of outage titles starts with ``Outage for''.
This pattern is easy for LLM to learn, so the titles generated by LLM tend to have more overlapping words, and consequently, higher scores.
Secondly, the words used in titles are usually either in a dictionary (\eg the ``Triage'' in the example title), or have been mentioned in incidents (\eg the ``Email Service'' in the example title).

Another observation is that title generation is the only task where IR outperforms the Rule-based method.
Since the Rule-based method performs simple concatenation, the generated title is long and contains unnecessary words, thus resulting in lower scores, while IR method retrieves titles from historical outages which conforms better with the pattern of outage titles in general.
\name achieves significantly high scores, with considerable improvement over baselines (at least 30.0\% of BLEU-4, 30.9\% of ROUGE-L, 31.4\% of METEOR), indicating that applying \name to production outages title generation is very promising.

\begin{table}[t]
\caption{
    Effectiveness of models at proposing outage titles
}
\begin{tabular}{lllllll}
\toprule
\multicolumn{1}{c}{\multirow{2}{*}{Model}} & \multicolumn{2}{c}{BLEU-4}        & \multicolumn{2}{c}{ROUGE-L}     & \multicolumn{2}{c}{METEOR}     \\
\multicolumn{1}{c}{}                       & Top1           & Top5           & Top1           & Top5           & Top1          & Top5           \\ \hline
IR                                         & 0.170          & 0.211          & 0.398          & 0.427          & 0.342         & 0.369          \\
Rule-based                                 & 0.069          & NA             & 0.211          & NA             & 0.316         & NA             \\
GPT-2                                       & 0.624          & 0.673          & 0.672          & 0.694          & 0.639         & 0.688          \\
Curie                                      & 0.826          & 0.88           & 0.88           & 0.9            & 0.84          & 0.894          \\
DaVinci & \textbf{0.857} & \textbf{0.893} & \textbf{0.883} & \textbf{0.913} & \textbf{0.869} & \textbf{0.907} \\ \bottomrule
\end{tabular}
\label{tab:exp2}
\vspacetbdown
\end{table}

\subsection{RQ6: Performance of Summary Generation Given Title}
We evaluate the performance of the outage summary generation when the title of the outage is given.
Remember that we provide the information of relevant incidents to these methods as context.
In this experiment, we include the title written by OCEs as part of the context.
For GPT-2 and \name where there is an instruction in the context ([\textit{The outage summary is:}]), we insert the outage title between incident information and the prompt.

The results of this experiment are listed in \tabref{tab:exp3}.
Surprisingly, the performance of \name, GPT-2, and Rule-based degrade slightly in this setting, with the exception of IR, whose metrics remain as low as before.
The relative order of methods in \tabref{tab:exp3} keeps the same as \tabref{tab:exp1}, for their tasks are very similar.
The performance degradation is because the title of the outage is not a grammatically-complete sentence.

\subsection{RQ7: Efficiency Comparison}

\tabref{tab:exp-time} lists the fine-tuning (if any) and summary generation times for each model. 
The fine-tuning time reflects the total time spent on all outages from the training set, while the summary generation time is the average time taken to generate a summary for one outage.
Despite the differences in parameter size, all models have a very small generation time. 
The fine-tuning time for LLMs (GPT-2, Curie, DaVinci) increases as the parameter size increases, although not linearly. 
Please note that \name only needs to be fine-tuned once on historical outages and incidents and can then be used for outage summary generation. 
In other words, the time to generate an outage summary of \name is only 13.3 or 39.6 ($\times10^{-5}$) time units, 
which is at least 251.2 times faster than the median of manual summarization.
In conclusion, the fine-tuning time for Oasis is reasonable, and its short generation time of summary demonstrates its practicality in cloud systems.

\begin{table}[t]
\caption{
    Effectiveness of  models at generating outage summaries given outage titles
}
\begin{tabular}{lllllll}
\toprule
\multicolumn{1}{c}{\multirow{2}{*}{Model}} & \multicolumn{2}{c}{BLEU-4}        & \multicolumn{2}{c}{ROUGE-L}     & \multicolumn{2}{c}{METEOR}     \\
\multicolumn{1}{c}{}                       & Top1           & Top5           & Top1           & Top5           & Top1           & Top5          \\ \hline
IR                                         & 0.037          & 0.055          & 0.156          & 0.189          & 0.109          & 0.142         \\
Rule-based                                 & 0.247          & NA             & 0.505          & NA             & 0.614          & NA            \\
GPT-2                                       & 0.428          & 0.504          & 0.548          & 0.59           & 0.515          & 0.569         \\
Curie                                      & 0.65           & 0.697          & 0.729          & 0.776          & 0.719          & 0.764         \\
DaVinci                                    & \textbf{0.652} & \textbf{0.699} & \textbf{0.734} & \textbf{0.779} & \textbf{0.724} & \textbf{0.77} \\ \bottomrule
\end{tabular}
\label{tab:exp3}
\vspacetbdown
\end{table}

\begin{table}[t]
\caption{Average time cost of models}
\begin{tabular}{llllll}
\toprule
Time       & Rule & IR & GPT-2 & Curie & DaVinci \\ \hline
\makecell[l]{
    Fine-tuning \\ ($10^{-1} \times $time unit)
    }  
& NA          & NA  &   3.0   &    3.4   &  8.7       \\
\makecell[l]{
    Generation \\ ($10^{-5} \times $ time unit)
    } &  2.8       & 13.9   & 11.1  &   13.3   &   39.6   \\ \bottomrule
\end{tabular}
\label{tab:exp-time}
\vspacetbdown
\end{table}

\section{OASIS Preliminary Human Evaluation}
\label{sec:human}

\subsection{Methodology}
Summarizing an outage is a challenging task that requires a deep understanding of both the service and the specific outage, as well as a comprehensive knowledge of the relevant context and domain.
To ensure the accuracy of the generated summaries, we ask the owners of these outages to evaluate the generated summaries from RQ4.
The process is done through Email and eventually, a total of 54 outage owners respond to our email and provide their evaluations.

For each outage, we present the generated summaries from all the methods, with the output of \name with Curie and DaVinci treated as independent summaries.
This results in five summaries for each outage.
We present the outage and the summaries in the following order:
(1) first we give the IcM link to the outage so that the outage owner can better recall the details of the outage,
(2) next \new{we present the original outage summary written by OCEs and tell the outage owners this is the human-written summarization}, 
(3) then we present the five outage summaries generated by models (\name-Curie, \name-DaVinci, GPT-2, Rule-based, IR), and we shuffle the order of these summaries to minimize the effect of default ordering.
To ensure the objectivity of the evaluation and avoid the subjectivity and bias of scoring for each summary, we ask the outage owners to rank the summaries from 1 to 5, where 1 for the most useful and readable and 5 for the least useful and readable.
\textit{Useful} means that the summary contains useful and relevant information on the outage.
\textit{Readable} means the ease with which the summary can be understood, which may be characterized by clear and simple language, logical organization, grammatical correctness, \etc
Besides ranking, we also ask outage owners to share their opinions and comments on model-generated summaries.

\subsection{Results}

\begin{figure}
    \centering
    \includegraphics[width=.76\linewidth]{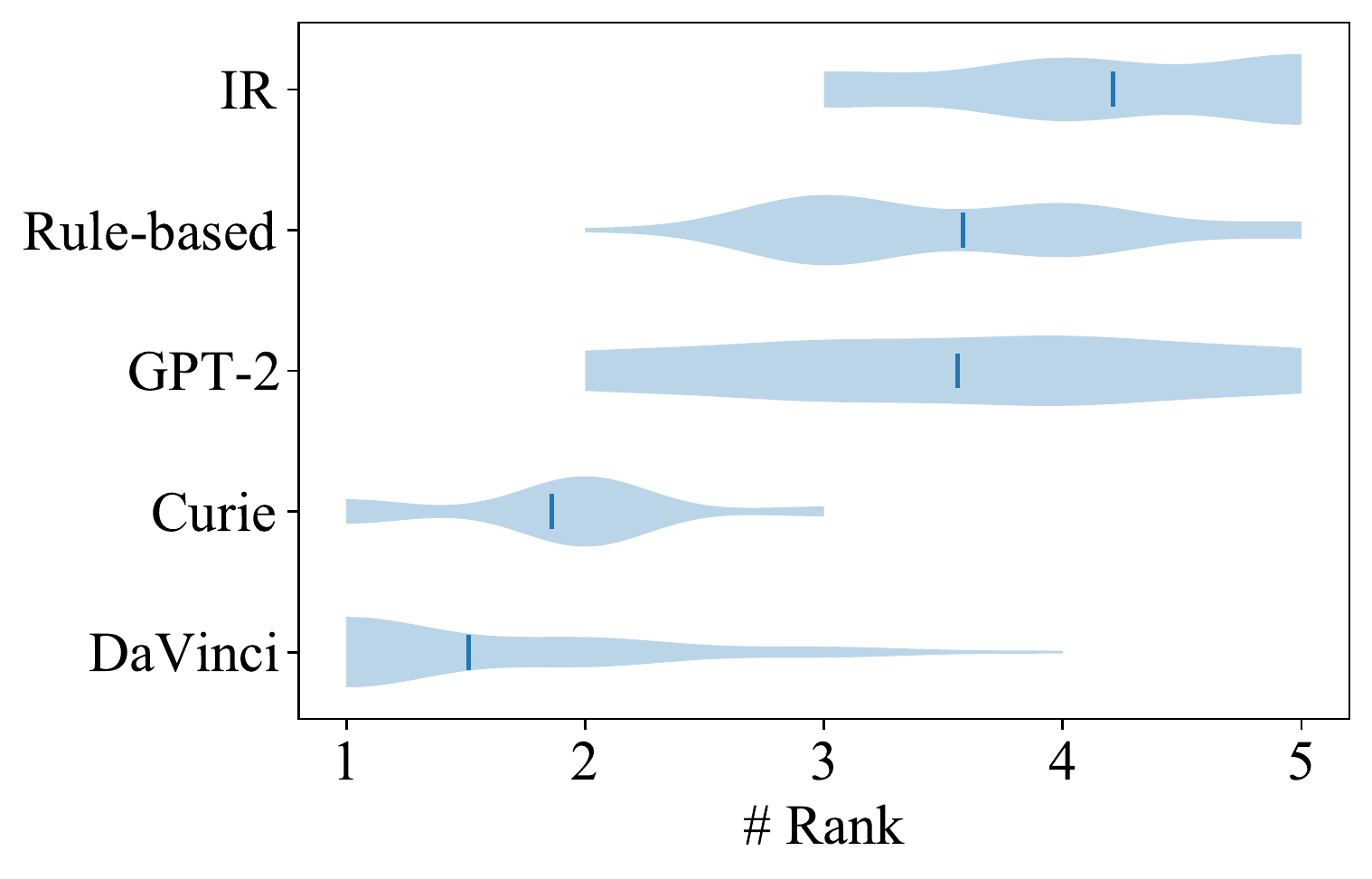}
    \caption{The ranking given by outage owners. Rank \#1 means the most preferred summary.}
    \label{fig:rank}
   
    \vspacefigdown
\end{figure}

\figref{fig:rank} shows the ranking of outage owners.
In general, the results of human evaluation are in accordance with automatic metrics presented in \tabref{tab:exp1}.
The outage owners report positive feedback regarding the readability and usefulness of \name.
\new{Notably, 32 out of 54 OCEs rank the summaries produced by \name-DaVinci as their top preference.}
To investigate whether the rankings of outage owners are consistent with each other,
we conduct Friedman Test \cite{eisinga2017exact} at the significant level of 0.05. 
The null hypothesis is that there is no significant difference between the rankings of outage owners. %
The calculated p-value on our rank by outage owner is larger than the level of significance, which means that the outage owners basically conform to each other in evaluating the summaries.

More encouragingly, the majority of outage owners have a favorable attitude toward the practice of generating outage summaries using \name:
\textit{\textbf{``I absolutely believe in the ability of AI to assist with incident management and outage summaries.
''}}

\section{Discussion}
\label{sec:discussions}

\textbf{Case Study} 

\noindent As described in \secref{subsec:imp}, \name serves as a supportive tool in the production IcM, \ie OCEs have the option to use or not use \name to generate a reference summary when handling outages. 
Since it is difficult to determine the extent to which \name contributes to the generated summary for outages where it is utilized, we randomly selected an outage from recent outages that were handled without the use of \name.

In this outage, a misconfiguration of the load balancer led to an overwhelming number of requests being directed to a single service endpoint. As a result, this endpoint was unable to function properly, causing the unified account API to fail. This failure cascaded to the downstream account APIs of the Cloud, Productivity, and Partner systems. As a result, the signup, ordering, and billing services of the Cloud and Productivity systems were affected.
Below is the actual outage summary, written by an experienced OCE:

\begin{center}
\begin{tcolorbox}[colback=gray!10,
  colframe=gray!10,
  width=\boxwidth,
  arc=2mm, auto outer arc,
  boxrule=0.5pt,
  left=\innerwidth,
  right=\innerwidth,
]      
\textbf{Outage Summary by OCE:~}{ 
The API$_{account}$ failed with HTTP 5xx errors (over $\alpha_1$ fall failures) because of bad gateway errors to the endpoint$_1$. Due to this issue, commercial customers could not sign-up for System-Cloud or System-Productivity via endpoint$_2$ or endpoint$_3$, and perform management related actions on endpoint$_4$. Additionally, customers could not complete purchases within these ecosystems. 
Partner system is also impacted.
} 
\end{tcolorbox}
\end{center}

\begin{center}
\begin{tcolorbox}[colback=gray!10,
  colframe=gray!10,
  width=\boxwidth,
  arc=2mm, auto outer arc,
  boxrule=0.5pt,
  left=\innerwidth,
  right=\innerwidth,
]
\textbf{Outage Summary by \name:~}{ 
The API$_{account}$ failed with HTTP 5xx errors (over $\alpha_1$ fall failures) because of bad gateway errors to the endpoint$_1$. Due to this issue, commercial customers could not sign-up for System-Cloud or System-Productivity via endpoint$_2$ or endpoint$_3$, and perform management related actions on endpoint$_4$. 
Additionally, System-Cloud users were not able to access their billing accounts and invoices on System-Cloud portal. Approximately $\alpha_2$ unique users were impacted.

} 
\end{tcolorbox}
\end{center}

To study the ability of \name in summarizing online outages, we triggered \name manually, limiting it to only knowing the information at the time of the outage.
\name managed to find six relevant incidents, and the generated summary is presented above.
(We indicate sentences that diverge from the OCE outage summary by underlining them with wavy lines)
In the above summaries, endpoints 1-4 are URLs that serve the API calls.
We notice that \name failed to identify System-Partner as impacted because the impact of System-Partner can only be determined by knowing the prefix of endpoint$_4$ refers to System-Partner.
This knowledge is difficult to learn even after the LLM has been fine-tuned using incident and outage corpus.
Despite this, we can see from the above example that \name is capable of generating reference summaries for outages. 
In this study, we adopt fine-tuning of the GPT-x models to generate outage summaries, which perform much better than prompt tuning according to our experiment.
Because the outage summary is very domain-specific and fine-tuned models may capture the domain knowledge.

\textbf{Threats to Validity}

\noindent Threats to \textbf{internal validity } mainly lie in our implementation of \name and compared approaches.
To reduce this threat, we implemented these approaches based on well-established frameworks, which have been described in \secref{sec:experiment-environment}.
Additionally, two authors carefully examined the code and configurations.

Threats to \textbf{external validity}  mainly lie in the subjects used in our study.
Our study and evaluation are conducted on 18 major cloud systems of \cloud.
Since the incidents and outages we used are only from \cloud, modifications may be necessary when applying to other incident management systems.
However, the cloud systems we used in our experiments include a variety of types, such as infrastructure, productivity, communication, game, search engine, \etc
Moreover, these cloud systems serve millions of customers, thus having a certain degree of representativeness.
In the future, we plan to extend our evaluations to include more cloud systems.

Threats to the \textbf{construct validity} mainly lie in the evaluation metrics we adopted.
Automated evaluation metrics (BLEU-4, ROUGE-L, and METEOR) may not fully reflect the readability and usefulness of the outage summary. To address this limitation, we will consider using additional metrics in the future to better measure these factors.
Moreover, we reached out to the owners of outages to conduct a human evaluation, and the evaluation results are basically aligned with automated metrics.

\section{Related Work}
\label{sec:related}
\point{Incident storm / outage handling}
Handling outages (incident storms) in cloud systems has been widely studied in previous work \cite{chen2020identifying, gu2020efficient, chen2021graph, wang2021fast}.
A series of works perform incident linking to provide engineers with more relevant information.
LiDAR \cite{chen2020identifying} calculates both textual similarity and component similarity to determine whether two incidents should be linked.
LinkCM \cite{gu2020efficient} argues that linking customer incidents (reported by customers) with system incidents (reported by monitors) can lead to more efficient incident triage.
The above works utilize neural networks to learn from historical linking patterns.
GRLIA \cite{chen2021graph} also employs graph embedding, with additional concerns about node closeness with KPI trend similarities.
COT \cite{wang2021fast} first builds a heuristic dependency graph for the cloud systems based on historical incidents and links. It then finds the corresponding incidents by searching connected nodes in the graph.
Another series of works focuses on alert reduction or prioritization \cite{chen2022online, zhao2020understanding}.
OAS \cite{chen2022online} combines semantic and behavioral features of alerts to decide the groups of alerts and then correlate alerts within a time window.
Zhao \etal \cite{zhao2020understanding} first calculate the textual and topological similarity of alerts to reduce the number of alerts. 
They then use DBSCAN to group similar alerts and selected the centroid alert of each cluster as the representative incident to show to engineers.
Our impact scope assessment is similar to these approaches, and we also include domain-specific knowledge via rule-based incident linking.

\point{Large Language Models (LLM) for Software Engineering}
In recent years, the rise of LLM has brought new opportunities to the field of software engineering \cite{mastropaolo2021studying, mastropaolo2022using, fu2022vulrepair, ahmed2023recommending, zhang2022using}.
Mastropaolo \etal \cite{mastropaolo2021studying} studied the ability of fine-tuned T5 in the following tasks: automatic bug fixing, generation of assert statements in test methods, code summarization, and injection of code mutants.
LANCE \cite{mastropaolo2022using} uses fine-tuned T5 to automatically generate logging statements for Java methods.
VulRepair \cite{fu2022vulrepair} also fine-tune T5 on vulnerability repairs datasets to automatically propose vulnerability fixes.
The above works fine-tune LLM on task-specific datasets.
Zhang \etal \cite{zhang2022using} propose to use prompting for LLM to improve code version control. 
They further integrate k-shot learning to resolve code merge conflicts.
GPT-3.x models are used to recommend root causes and mitigation steps to facilitate cloud incident management \cite{ahmed2023recommending}.
Different from previous studies, \name is the first work to leverage the capabilities of LLM in to summarize outages for cloud systems.

\section{Conclusion}
\label{sec:conclusion}

In this paper, we identify the problem of outage understanding in real-world cloud systems.
Through our empirical study on 18 industrial cloud systems, we show that understanding outage is time-consuming and involves complex contexts.
To improve the process of outage understanding, we present \name, the first framework to automatically assess impacts and summarize outages.
\name incorporates an assessment of outage impact scope and a fine-tuned large language model, \ie GPT-3.x.
Our experiments on 18 cloud systems within \cloud demonstrate that \name outperforms baseline approaches. 
We also received feedback from outage owners, which further validates the effectiveness of \name.

\bibliographystyle{ACM-Reference-Format}
\balance 
\bibliography{references}


\begin{thebibliography}{31}

%%% the Web address.

\ifx \showCODEN    \undefined \def \showCODEN     #1{\unskip}     \fi
\ifx \showDOI      \undefined \def \showDOI       #1{#1}\fi
\ifx \showISBNx    \undefined \def \showISBNx     #1{\unskip}     \fi
\ifx \showISBNxiii \undefined \def \showISBNxiii  #1{\unskip}     \fi
\ifx \showISSN     \undefined \def \showISSN      #1{\unskip}     \fi
\ifx \showLCCN     \undefined \def \showLCCN      #1{\unskip}     \fi
\ifx \shownote     \undefined \def \shownote      #1{#1}          \fi
\ifx \showarticletitle \undefined \def \showarticletitle #1{#1}   \fi
\ifx \showURL      \undefined \def \showURL       {\relax}        \fi
% invisible to TeX
\providecommand\bibfield[2]{#2}
\providecommand\bibinfo[2]{#2}
\providecommand\natexlab[1]{#1}
\providecommand\showeprint[2][]{arXiv:#2}

\bibitem[Ahmed and Devanbu(2022)]%
        {ahmed2022multilingual}
\bibfield{author}{\bibinfo{person}{Toufique Ahmed} {and}
  \bibinfo{person}{Premkumar Devanbu}.} \bibinfo{year}{2022}\natexlab{}.
\newblock \showarticletitle{Multilingual training for software engineering}. In
  \bibinfo{booktitle}{\emph{Proceedings of the 44th International Conference on
  Software Engineering}}. \bibinfo{pages}{1443--1455}.
\newblock


\bibitem[Ahmed et~al\mbox{.}(2023)]%
        {ahmed2023recommending}
\bibfield{author}{\bibinfo{person}{Toufique Ahmed}, \bibinfo{person}{Supriyo
  Ghosh}, \bibinfo{person}{Chetan Bansal}, \bibinfo{person}{Thomas Zimmermann},
  \bibinfo{person}{Xuchao Zhang}, {and} \bibinfo{person}{Saravan Rajmohan}.}
  \bibinfo{year}{2023}\natexlab{}.
\newblock \showarticletitle{Recommending Root-Cause and Mitigation Steps for
  Cloud Incidents using Large Language Models}. In
  \bibinfo{booktitle}{\emph{ICSE 2023}}.
\newblock


\bibitem[Brown et~al\mbox{.}(2020)]%
        {brown2020language}
\bibfield{author}{\bibinfo{person}{Tom Brown}, \bibinfo{person}{Benjamin Mann},
  \bibinfo{person}{Nick Ryder}, \bibinfo{person}{Melanie Subbiah},
  \bibinfo{person}{Jared~D Kaplan}, \bibinfo{person}{Prafulla Dhariwal},
  \bibinfo{person}{Arvind Neelakantan}, \bibinfo{person}{Pranav Shyam},
  \bibinfo{person}{Girish Sastry}, \bibinfo{person}{Amanda Askell},
  {et~al\mbox{.}}} \bibinfo{year}{2020}\natexlab{}.
\newblock \showarticletitle{Language models are few-shot learners}.
\newblock \bibinfo{journal}{\emph{Advances in neural information processing
  systems}}  \bibinfo{volume}{33} (\bibinfo{year}{2020}),
  \bibinfo{pages}{1877--1901}.
\newblock


\bibitem[Chen et~al\mbox{.}(2019a)]%
        {chen2019empirical}
\bibfield{author}{\bibinfo{person}{Junjie Chen}, \bibinfo{person}{Xiaoting He},
  \bibinfo{person}{Qingwei Lin}, \bibinfo{person}{Yong Xu},
  \bibinfo{person}{Hongyu Zhang}, \bibinfo{person}{Dan Hao},
  \bibinfo{person}{Feng Gao}, \bibinfo{person}{Zhangwei Xu},
  \bibinfo{person}{Yingnong Dang}, {and} \bibinfo{person}{Dongmei Zhang}.}
  \bibinfo{year}{2019}\natexlab{a}.
\newblock \showarticletitle{An empirical investigation of incident triage for
  online service systems}. In \bibinfo{booktitle}{\emph{IEEE/ACM 41st
  International Conference on Software Engineering: Software Engineering in
  Practice (ICSE-SEIP)}}. IEEE, \bibinfo{pages}{111--120}.
\newblock


\bibitem[Chen et~al\mbox{.}(2019b)]%
        {chen2019continuous}
\bibfield{author}{\bibinfo{person}{Junjie Chen}, \bibinfo{person}{Xiaoting He},
  \bibinfo{person}{Qingwei Lin}, \bibinfo{person}{Hongyu Zhang},
  \bibinfo{person}{Dan Hao}, \bibinfo{person}{Feng Gao},
  \bibinfo{person}{Zhangwei Xu}, \bibinfo{person}{Yingnong Dang}, {and}
  \bibinfo{person}{Dongmei Zhang}.} \bibinfo{year}{2019}\natexlab{b}.
\newblock \showarticletitle{Continuous incident triage for large-scale online
  service systems}. In \bibinfo{booktitle}{\emph{2019 34th IEEE/ACM
  International Conference on Automated Software Engineering (ASE)}}. IEEE,
  \bibinfo{pages}{364--375}.
\newblock


\bibitem[Chen et~al\mbox{.}(2022)]%
        {chen2022online}
\bibfield{author}{\bibinfo{person}{Jia Chen}, \bibinfo{person}{Peng Wang},
  {and} \bibinfo{person}{Wei Wang}.} \bibinfo{year}{2022}\natexlab{}.
\newblock \showarticletitle{Online Summarizing Alerts through Semantic and
  Behavior Information}. In \bibinfo{booktitle}{\emph{Proceedings of the 44th
  International Conference on Software Engineering}} (Pittsburgh, Pennsylvania)
  \emph{(\bibinfo{series}{ICSE '22})}. \bibinfo{pages}{1646–1657}.
\newblock
\urldef\tempurl%
\url{https://doi.org/10.1145/3510003.3510055}
\showDOI{\tempurl}


\bibitem[Chen et~al\mbox{.}(2020c)]%
        {chen2020incidental}
\bibfield{author}{\bibinfo{person}{Junjie Chen}, \bibinfo{person}{Shu Zhang},
  \bibinfo{person}{Xiaoting He}, \bibinfo{person}{Qingwei Lin},
  \bibinfo{person}{Hongyu Zhang}, \bibinfo{person}{Dan Hao},
  \bibinfo{person}{Yu Kang}, \bibinfo{person}{Feng Gao},
  \bibinfo{person}{Zhangwei Xu}, \bibinfo{person}{Yingnong Dang},
  {et~al\mbox{.}}} \bibinfo{year}{2020}\natexlab{c}.
\newblock \showarticletitle{How incidental are the incidents? characterizing
  and prioritizing incidents for large-scale online service systems}. In
  \bibinfo{booktitle}{\emph{Proceedings of the 35th IEEE/ACM International
  Conference on Automated Software Engineering}}. \bibinfo{pages}{373--384}.
\newblock


\bibitem[Chen et~al\mbox{.}(2020b)]%
        {chen2020identifying}
\bibfield{author}{\bibinfo{person}{Yujun Chen}, \bibinfo{person}{Xian Yang},
  \bibinfo{person}{Hang Dong}, \bibinfo{person}{Xiaoting He},
  \bibinfo{person}{Hongyu Zhang}, \bibinfo{person}{Qingwei Lin},
  \bibinfo{person}{Junjie Chen}, \bibinfo{person}{Pu Zhao}, \bibinfo{person}{Yu
  Kang}, \bibinfo{person}{Feng Gao}, {et~al\mbox{.}}}
  \bibinfo{year}{2020}\natexlab{b}.
\newblock \showarticletitle{Identifying linked incidents in large-scale online
  service systems}. In \bibinfo{booktitle}{\emph{Proceedings of the 28th ACM
  Joint Meeting on European Software Engineering Conference and Symposium on
  the Foundations of Software Engineering}}. \bibinfo{pages}{304--314}.
\newblock


\bibitem[Chen et~al\mbox{.}(2019c)]%
        {chen2019outage}
\bibfield{author}{\bibinfo{person}{Yujun Chen}, \bibinfo{person}{Xian Yang},
  \bibinfo{person}{Qingwei Lin}, \bibinfo{person}{Hongyu Zhang},
  \bibinfo{person}{Feng Gao}, \bibinfo{person}{Zhangwei Xu},
  \bibinfo{person}{Yingnong Dang}, \bibinfo{person}{Dongmei Zhang},
  \bibinfo{person}{Hang Dong}, \bibinfo{person}{Yong Xu}, {et~al\mbox{.}}}
  \bibinfo{year}{2019}\natexlab{c}.
\newblock \showarticletitle{Outage prediction and diagnosis for cloud service
  systems}. In \bibinfo{booktitle}{\emph{The world wide web conference}}.
  \bibinfo{pages}{2659--2665}.
\newblock


\bibitem[Chen et~al\mbox{.}(2020a)]%
        {chen2020towards}
\bibfield{author}{\bibinfo{person}{Zhuangbin Chen}, \bibinfo{person}{Yu Kang},
  \bibinfo{person}{Liqun Li}, \bibinfo{person}{Xu Zhang},
  \bibinfo{person}{Hongyu Zhang}, \bibinfo{person}{Hui Xu},
  \bibinfo{person}{Yangfan Zhou}, \bibinfo{person}{Li Yang},
  \bibinfo{person}{Jeffrey Sun}, \bibinfo{person}{Zhangwei Xu},
  {et~al\mbox{.}}} \bibinfo{year}{2020}\natexlab{a}.
\newblock \showarticletitle{Towards intelligent incident management: why we
  need it and how we make it}. In \bibinfo{booktitle}{\emph{Proceedings of the
  28th ACM Joint Meeting on European Software Engineering Conference and
  Symposium on the Foundations of Software Engineering}}.
  \bibinfo{pages}{1487--1497}.
\newblock


\bibitem[Chen et~al\mbox{.}(2021)]%
        {chen2021graph}
\bibfield{author}{\bibinfo{person}{Zhuangbin Chen}, \bibinfo{person}{Jinyang
  Liu}, \bibinfo{person}{Yuxin Su}, \bibinfo{person}{Hongyu Zhang},
  \bibinfo{person}{Xuemin Wen}, \bibinfo{person}{Xiao Ling},
  \bibinfo{person}{Yongqiang Yang}, {and} \bibinfo{person}{Michael~R Lyu}.}
  \bibinfo{year}{2021}\natexlab{}.
\newblock \showarticletitle{Graph-based incident aggregation for large-scale
  online service systems}. In \bibinfo{booktitle}{\emph{2021 36th IEEE/ACM
  International Conference on Automated Software Engineering (ASE)}}. IEEE,
  \bibinfo{pages}{430--442}.
\newblock


\bibitem[Eisinga et~al\mbox{.}(2017)]%
        {eisinga2017exact}
\bibfield{author}{\bibinfo{person}{Rob Eisinga}, \bibinfo{person}{Tom Heskes},
  \bibinfo{person}{Ben Pelzer}, {and} \bibinfo{person}{Manfred Te~Grotenhuis}.}
  \bibinfo{year}{2017}\natexlab{}.
\newblock \showarticletitle{Exact p-values for pairwise comparison of Friedman
  rank sums, with application to comparing classifiers}.
\newblock \bibinfo{journal}{\emph{BMC bioinformatics}} \bibinfo{volume}{18},
  \bibinfo{number}{1} (\bibinfo{year}{2017}), \bibinfo{pages}{1--18}.
\newblock


\bibitem[Fu et~al\mbox{.}(2022)]%
        {fu2022vulrepair}
\bibfield{author}{\bibinfo{person}{Michael Fu}, \bibinfo{person}{Chakkrit
  Tantithamthavorn}, \bibinfo{person}{Trung Le}, \bibinfo{person}{Van Nguyen},
  {and} \bibinfo{person}{Dinh Phung}.} \bibinfo{year}{2022}\natexlab{}.
\newblock \showarticletitle{VulRepair: a T5-based automated software
  vulnerability repair}. In \bibinfo{booktitle}{\emph{Proceedings of the 30th
  ACM Joint European Software Engineering Conference and Symposium on the
  Foundations of Software Engineering}}. \bibinfo{pages}{935--947}.
\newblock


\bibitem[Ghosh et~al\mbox{.}(2022)]%
        {ghosh2022fight}
\bibfield{author}{\bibinfo{person}{Supriyo Ghosh}, \bibinfo{person}{Manish
  Shetty}, \bibinfo{person}{Chetan Bansal}, {and} \bibinfo{person}{Suman
  Nath}.} \bibinfo{year}{2022}\natexlab{}.
\newblock \showarticletitle{How to fight production incidents? an empirical
  study on a large-scale cloud service}. In
  \bibinfo{booktitle}{\emph{Proceedings of the 13th Symposium on Cloud
  Computing}}. \bibinfo{pages}{126--141}.
\newblock


\bibitem[Gros et~al\mbox{.}(2020)]%
        {gros2020code}
\bibfield{author}{\bibinfo{person}{David Gros}, \bibinfo{person}{Hariharan
  Sezhiyan}, \bibinfo{person}{Prem Devanbu}, {and} \bibinfo{person}{Zhou Yu}.}
  \bibinfo{year}{2020}\natexlab{}.
\newblock \showarticletitle{Code to comment" translation" data, metrics,
  baselining \& evaluation}. In \bibinfo{booktitle}{\emph{Proceedings of the
  35th IEEE/ACM International Conference on Automated Software Engineering}}.
  \bibinfo{pages}{746--757}.
\newblock


\bibitem[Gu et~al\mbox{.}(2020a)]%
        {gu2020efficientincident}
\bibfield{author}{\bibinfo{person}{Jiazhen Gu}, \bibinfo{person}{Chuan Luo},
  \bibinfo{person}{Si Qin}, \bibinfo{person}{Bo Qiao}, \bibinfo{person}{Qingwei
  Lin}, \bibinfo{person}{Hongyu Zhang}, \bibinfo{person}{Ze Li},
  \bibinfo{person}{Yingnong Dang}, \bibinfo{person}{Shaowei Cai},
  \bibinfo{person}{Wei Wu}, {et~al\mbox{.}}} \bibinfo{year}{2020}\natexlab{a}.
\newblock \showarticletitle{Efficient incident identification from
  multi-dimensional issue reports via meta-heuristic search}. In
  \bibinfo{booktitle}{\emph{Proceedings of the 28th ACM Joint Meeting on
  European Software Engineering Conference and Symposium on the Foundations of
  Software Engineering}}. \bibinfo{pages}{292--303}.
\newblock


\bibitem[Gu et~al\mbox{.}(2020b)]%
        {gu2020efficient}
\bibfield{author}{\bibinfo{person}{Jiazhen Gu}, \bibinfo{person}{Jiaqi Wen},
  \bibinfo{person}{Zijian Wang}, \bibinfo{person}{Pu Zhao},
  \bibinfo{person}{Chuan Luo}, \bibinfo{person}{Yu Kang},
  \bibinfo{person}{Yangfan Zhou}, \bibinfo{person}{Li Yang},
  \bibinfo{person}{Jeffrey Sun}, \bibinfo{person}{Zhangwei Xu},
  {et~al\mbox{.}}} \bibinfo{year}{2020}\natexlab{b}.
\newblock \showarticletitle{Efficient customer incident triage via linking with
  system incidents}. In \bibinfo{booktitle}{\emph{Proceedings of the 28th ACM
  Joint Meeting on European Software Engineering Conference and Symposium on
  the Foundations of Software Engineering}}. \bibinfo{pages}{1296--1307}.
\newblock


\bibitem[Hadary et~al\mbox{.}(2020)]%
        {ori2020protean}
\bibfield{author}{\bibinfo{person}{Ori Hadary}, \bibinfo{person}{Luke
  Marshall}, \bibinfo{person}{Ishai Menache}, \bibinfo{person}{Abhisek Pan},
  \bibinfo{person}{Esaias~E. Greeff}, \bibinfo{person}{David Dion},
  \bibinfo{person}{Star Dorminey}, \bibinfo{person}{Shailesh Joshi},
  \bibinfo{person}{Yang Chen}, \bibinfo{person}{Mark Russinovich}, {and}
  \bibinfo{person}{Thomas Moscibroda}.} \bibinfo{year}{2020}\natexlab{}.
\newblock \showarticletitle{Protean: {VM} Allocation Service at Scale}. In
  \bibinfo{booktitle}{\emph{14th {USENIX} Symposium on Operating Systems Design
  and Implementation, {OSDI} 2020, Virtual Event, November 4-6, 2020}}.
  \bibinfo{publisher}{{USENIX} Association}, \bibinfo{pages}{845--861}.
\newblock
\urldef\tempurl%
\url{https://www.usenix.org/conference/osdi20/presentation/hadary}
\showURL{%
\tempurl}


\bibitem[Jiang et~al\mbox{.}(2020)]%
        {jiang2020mitigate}
\bibfield{author}{\bibinfo{person}{Jiajun Jiang}, \bibinfo{person}{Weihai Lu},
  \bibinfo{person}{Junjie Chen}, \bibinfo{person}{Qingwei Lin},
  \bibinfo{person}{Pu Zhao}, \bibinfo{person}{Yu Kang}, \bibinfo{person}{Hongyu
  Zhang}, \bibinfo{person}{Yingfei Xiong}, \bibinfo{person}{Feng Gao},
  \bibinfo{person}{Zhangwei Xu}, {et~al\mbox{.}}}
  \bibinfo{year}{2020}\natexlab{}.
\newblock \showarticletitle{How to mitigate the incident? an effective
  troubleshooting guide recommendation technique for online service systems}.
  In \bibinfo{booktitle}{\emph{Proceedings of the 28th ACM Joint Meeting on
  European Software Engineering Conference and Symposium on the Foundations of
  Software Engineering}}. \bibinfo{pages}{1410--1420}.
\newblock


\bibitem[Li et~al\mbox{.}(2021)]%
        {li2021fighting}
\bibfield{author}{\bibinfo{person}{Liqun Li}, \bibinfo{person}{Xu Zhang},
  \bibinfo{person}{Xin Zhao}, \bibinfo{person}{Hongyu Zhang},
  \bibinfo{person}{Yu Kang}, \bibinfo{person}{Pu Zhao}, \bibinfo{person}{Bo
  Qiao}, \bibinfo{person}{Shilin He}, \bibinfo{person}{Pochian Lee},
  \bibinfo{person}{Jeffrey Sun}, {et~al\mbox{.}}}
  \bibinfo{year}{2021}\natexlab{}.
\newblock \showarticletitle{Fighting the Fog of War: Automated Incident
  Detection for Cloud Systems}. In \bibinfo{booktitle}{\emph{USENIX Annual
  Technical Conference}}. \bibinfo{pages}{131--146}.
\newblock


\bibitem[Liu et~al\mbox{.}(2019)]%
        {liu2019bugs}
\bibfield{author}{\bibinfo{person}{Haopeng Liu}, \bibinfo{person}{Shan Lu},
  \bibinfo{person}{Madan Musuvathi}, {and} \bibinfo{person}{Suman Nath}.}
  \bibinfo{year}{2019}\natexlab{}.
\newblock \showarticletitle{What bugs cause production cloud incidents?}. In
  \bibinfo{booktitle}{\emph{Proceedings of the Workshop on Hot Topics in
  Operating Systems}}. \bibinfo{pages}{155--162}.
\newblock


\bibitem[Liu et~al\mbox{.}(2018)]%
        {liu2018neural}
\bibfield{author}{\bibinfo{person}{Zhongxin Liu}, \bibinfo{person}{Xin Xia},
  \bibinfo{person}{Ahmed~E Hassan}, \bibinfo{person}{David Lo},
  \bibinfo{person}{Zhenchang Xing}, {and} \bibinfo{person}{Xinyu Wang}.}
  \bibinfo{year}{2018}\natexlab{}.
\newblock \showarticletitle{Neural-machine-translation-based commit message
  generation: how far are we?}. In \bibinfo{booktitle}{\emph{Proceedings of the
  33rd ACM/IEEE International Conference on Automated Software Engineering}}.
  \bibinfo{pages}{373--384}.
\newblock


\bibitem[Ma et~al\mbox{.}(2022)]%
        {ma2022empirical}
\bibfield{author}{\bibinfo{person}{Minghua Ma}, \bibinfo{person}{Yudong Liu},
  \bibinfo{person}{Yuang Tong}, \bibinfo{person}{Haozhe Li},
  \bibinfo{person}{Pu Zhao}, \bibinfo{person}{Yong Xu}, \bibinfo{person}{Hongyu
  Zhang}, \bibinfo{person}{Shilin He}, \bibinfo{person}{Lu Wang},
  \bibinfo{person}{Yingnong Dang}, \bibinfo{person}{Saravanakumar Rajmohan},
  {and} \bibinfo{person}{Qingwei Lin}.} \bibinfo{year}{2022}\natexlab{}.
\newblock \showarticletitle{An Empirical Investigation of Missing Data Handling
  in Cloud Node Failure Prediction}. In \bibinfo{booktitle}{\emph{Proceedings
  of the European Software Engineering Conference and Symposium on the
  Foundations of Software Engineering (ESEC/FSE)}}. \bibinfo{pages}{1453 --
  1464}.
\newblock


\bibitem[Ma et~al\mbox{.}(2020)]%
        {ma2020diagnosing}
\bibfield{author}{\bibinfo{person}{Minghua Ma}, \bibinfo{person}{Zheng Yin},
  \bibinfo{person}{Shenglin Zhang}, \bibinfo{person}{Sheng Wang},
  \bibinfo{person}{Christopher Zheng}, \bibinfo{person}{Xinhao Jiang},
  \bibinfo{person}{Hanwen Hu}, \bibinfo{person}{Cheng Luo},
  \bibinfo{person}{Yilin Li}, \bibinfo{person}{Nengjun Qiu}, {et~al\mbox{.}}}
  \bibinfo{year}{2020}\natexlab{}.
\newblock \showarticletitle{Diagnosing root causes of intermittent slow queries
  in cloud databases}.
\newblock \bibinfo{journal}{\emph{Proceedings of the VLDB Endowment}}
  \bibinfo{volume}{13}, \bibinfo{number}{8} (\bibinfo{year}{2020}),
  \bibinfo{pages}{1176--1189}.
\newblock


\bibitem[Mastropaolo et~al\mbox{.}(2022)]%
        {mastropaolo2022using}
\bibfield{author}{\bibinfo{person}{Antonio Mastropaolo}, \bibinfo{person}{Luca
  Pascarella}, {and} \bibinfo{person}{Gabriele Bavota}.}
  \bibinfo{year}{2022}\natexlab{}.
\newblock \showarticletitle{Using Deep Learning to Generate Complete Log
  Statements}. In \bibinfo{booktitle}{\emph{Proceedings of the 44th
  International Conference on Software Engineering}}
  \emph{(\bibinfo{series}{ICSE '22})}. \bibinfo{pages}{2279–2290}.
\newblock


\bibitem[Mastropaolo et~al\mbox{.}(2021)]%
        {mastropaolo2021studying}
\bibfield{author}{\bibinfo{person}{Antonio Mastropaolo},
  \bibinfo{person}{Simone Scalabrino}, \bibinfo{person}{Nathan Cooper},
  \bibinfo{person}{David~Nader Palacio}, \bibinfo{person}{Denys Poshyvanyk},
  \bibinfo{person}{Rocco Oliveto}, {and} \bibinfo{person}{Gabriele Bavota}.}
  \bibinfo{year}{2021}\natexlab{}.
\newblock \showarticletitle{Studying the usage of text-to-text transfer
  transformer to support code-related tasks}. In \bibinfo{booktitle}{\emph{2021
  IEEE/ACM 43rd International Conference on Software Engineering (ICSE)}}.
  IEEE, \bibinfo{pages}{336--347}.
\newblock


\bibitem[Radford et~al\mbox{.}(2019)]%
        {radford2019language}
\bibfield{author}{\bibinfo{person}{Alec Radford}, \bibinfo{person}{Jeffrey Wu},
  \bibinfo{person}{Rewon Child}, \bibinfo{person}{David Luan},
  \bibinfo{person}{Dario Amodei}, \bibinfo{person}{Ilya Sutskever},
  {et~al\mbox{.}}} \bibinfo{year}{2019}\natexlab{}.
\newblock \showarticletitle{Language models are unsupervised multitask
  learners}.
\newblock \bibinfo{journal}{\emph{OpenAI blog}} \bibinfo{volume}{1},
  \bibinfo{number}{8} (\bibinfo{year}{2019}), \bibinfo{pages}{9}.
\newblock


\bibitem[Vaswani et~al\mbox{.}(2017)]%
        {vaswani2017attention}
\bibfield{author}{\bibinfo{person}{Ashish Vaswani}, \bibinfo{person}{Noam
  Shazeer}, \bibinfo{person}{Niki Parmar}, \bibinfo{person}{Jakob Uszkoreit},
  \bibinfo{person}{Llion Jones}, \bibinfo{person}{Aidan~N Gomez},
  \bibinfo{person}{{\L}ukasz Kaiser}, {and} \bibinfo{person}{Illia
  Polosukhin}.} \bibinfo{year}{2017}\natexlab{}.
\newblock \showarticletitle{Attention is all you need}.
\newblock \bibinfo{journal}{\emph{Advances in neural information processing
  systems}}  \bibinfo{volume}{30} (\bibinfo{year}{2017}).
\newblock


\bibitem[Wang et~al\mbox{.}(2021)]%
        {wang2021fast}
\bibfield{author}{\bibinfo{person}{Yaohui Wang}, \bibinfo{person}{Guozheng Li},
  \bibinfo{person}{Zijian Wang}, \bibinfo{person}{Yu Kang},
  \bibinfo{person}{Yangfan Zhou}, \bibinfo{person}{Hongyu Zhang},
  \bibinfo{person}{Feng Gao}, \bibinfo{person}{Jeffrey Sun},
  \bibinfo{person}{Li Yang}, \bibinfo{person}{Pochian Lee}, {et~al\mbox{.}}}
  \bibinfo{year}{2021}\natexlab{}.
\newblock \showarticletitle{Fast outage analysis of large-scale production
  clouds with service correlation mining}. In \bibinfo{booktitle}{\emph{2021
  IEEE/ACM 43rd International Conference on Software Engineering (ICSE)}}.
  IEEE, \bibinfo{pages}{885--896}.
\newblock


\bibitem[Zhang et~al\mbox{.}(2022)]%
        {zhang2022using}
\bibfield{author}{\bibinfo{person}{Jialu Zhang}, \bibinfo{person}{Todd
  Mytkowicz}, \bibinfo{person}{Mike Kaufman}, \bibinfo{person}{Ruzica Piskac},
  {and} \bibinfo{person}{Shuvendu~K Lahiri}.} \bibinfo{year}{2022}\natexlab{}.
\newblock \showarticletitle{Using pre-trained language models to resolve
  textual and semantic merge conflicts (experience paper)}. In
  \bibinfo{booktitle}{\emph{Proceedings of the 31st ACM SIGSOFT International
  Symposium on Software Testing and Analysis}}. \bibinfo{pages}{77--88}.
\newblock


\bibitem[Zhao et~al\mbox{.}(2020)]%
        {zhao2020understanding}
\bibfield{author}{\bibinfo{person}{Nengwen Zhao}, \bibinfo{person}{Junjie
  Chen}, \bibinfo{person}{Xiao Peng}, \bibinfo{person}{Honglin Wang},
  \bibinfo{person}{Xinya Wu}, \bibinfo{person}{Yuanzong Zhang},
  \bibinfo{person}{Zikai Chen}, \bibinfo{person}{Xiangzhong Zheng},
  \bibinfo{person}{Xiaohui Nie}, \bibinfo{person}{Gang Wang}, {et~al\mbox{.}}}
  \bibinfo{year}{2020}\natexlab{}.
\newblock \showarticletitle{Understanding and handling alert storm for online
  service systems}. In \bibinfo{booktitle}{\emph{2020 IEEE/ACM 42nd
  International Conference on Software Engineering: Software Engineering in
  Practice (ICSE-SEIP)}}. IEEE, \bibinfo{pages}{162--171}.
\newblock


\end{thebibliography}

\end{document}